\documentclass[11pt]{article}
\renewcommand{\theequation}{\arabic{section}.\arabic{equation}}
\def\lb{\label}
\def\bb{\bibitem}
\def\be{\begin{equation}}
\def\ee{\end{equation}}
\def\ba{\begin{eqnarray}}
\def\ea{\end{eqnarray}}
\def\ds{\displaystyle}
\def\nn{\nonumber}
\def\tom{\tilde\omega}
\def\5{_{(5)}}
\begin{document}
\begin{titlepage}
\title{
\begin{flushright}\begin{small}
LAPTH-1209/07
\end{small}\end{flushright}
\vspace{2cm} The symmetries of five-dimensional minimal supergravity
reduced to three dimensions} \author{G\'erard Cl\'ement
\thanks{Email: gclement@lapp.in2p3.fr} \\ \\ \small Laboratoire de
Physique Th\'eorique LAPTH (CNRS), \\ \small B.P.110, F-74941
Annecy-le-Vieux cedex, France} \date{4 October 2007} \maketitle
\abstract{The 14 Killing vectors of the target space for
five-dimensional minimal supergravity reduced to three dimensions are
explicitly constructed in terms  of the original field
variables. These vectors generate the Lie algebra of $G_2$. We also
construct a symmetrical $7\times7$ matrix representative of the coset
$G_{2(+2)}/((SL(2,R)\times SL(2,R))$ as a function of the same fields.}

\end{titlepage}
\setcounter{page}{2}
\section{Introduction}

A number of self-gravitating field theories in $D$ dimensions can be
dimensionally reduced to gravity-coupled sigma models in three dimensions
\cite{BM84,CJLP}. Such sigma models are harmonic maps from a 3-dimensional
(Minkowskian or Lorentzian) base space to a $p$-dimensional target space
$\cal T$. The target space is generically a coset $G/H$, where $G$ is the group
of global isometries of $\cal T$, and $H \in G$ the local isotropy subgroup.
It is then possible to generate new solutions by applying a finite group
transformation to the coset representative of a seed solution
\cite{neukra,kin73}.
Another fruitful application is the construction of multi-center solutions as
totally geodesic submanifolds of the target space \cite{multi}.

The reduction of $D=11$ supergravity to $D=3$ leads to the
$E_{8(+8)}/SO(16)$ sigma model \cite{julia81,MS}. More recently, it
was shown  that the reduction of five-dimensional minimal
supergravity \cite{cremmer81,CN} leads to the $G_{2(+2)}/SO(4)$
sigma model \cite{mizoh,mizschro,CJLP,PS} in the case of a
Lorentzian 3-space, or $G_{2(+2)}/((SL(2,R)\times SL(2,R))$ for a
Minkowskian 3-space. Matrix representations of this coset were given
in these papers in terms of the 14-dimensional adjoint
representation. A representation of the same coset as a 7x7 matrix
was given in \cite{Nikolai}, however the parametrisation used leads
to a matrix which is too complicated to be used for solution
generation. The purpose of this paper is two-fold. First, we shall
give an analytical construction of the  14 infinitesimal isometries
(or Killing vectors) of the target space for dimensionally reduced
five-dimensional minimal supergravity, and check that they generate
the Lie algebra of  the exceptional group $G_2$ \footnote{These
Killing vectors were previously  determined, using a different
parametrisation, by a computer-assisted solution of the Killing
equations \cite{g2}.}. Then, we shall construct a coset
representative as a symmetrical $7\times7$ matrix.

Our reduction from five to three dimensions follows essentially the
same path as in \cite{mizoh}. With a view for the paper to be
self-contained, we outline  this reduction in Sect. 2. The result is
the metric (\ref{tarmet}) for an  eight-dimensional target space
$\cal T$. The procedure followed in \cite{mizoh} to identify $\cal
T$ as the $G_{2(+2)}/SO(4)$ coset was to construct a matrix
representative of this coset in the adjoint representation of $G_2$,
and to  show that the resulting metric was isometric to
(\ref{tarmet}). We shall instead follow a `bottom-up' approach. In
Sect. 3 we show that, taking into account the field-theoretical
construction of (\ref{tarmet}), we can identify nine manifest
infinitesimal symmetries of this metric. Combining this information
with the assumption that the unknown symmetry group $G$ must contain
$SL(3,R$), which is the isometry group for the vacuum sector
(five-dimensional Einstein gravity reduced to three
dimensions)\cite{maison}, we show that the minimal Lie algebra
necessarily closes to that of $G_2$. We solve in Sect. 4 the Lie
brackets involving the five unknown generators, and determine these
up to a single integration constant. We then determine the value of
this constant so that these five generators are indeed Killing
vectors of the metric (\ref{tarmet}). In Sect. 5, a symmetrical
$7\times7$ matrix coset representative is obtained by exponentiating
a Borel subalgebra. An alternative construction of the same matrix
using the nine manifest Killing vectors of Sect. 3 is sketched in
the Appendix.

\section{Five-to-three dimensional reduction}
\setcounter{equation}{0}

The bosonic sector of five-dimensional minimal supergravity is defined
by the Einstein-Maxwell-Chern-Simons action \ba\lb{em5} S_5 &=&
\frac1{16\pi G_5}\int d^5x \bigg[\sqrt{|g_{(5)}|} \bigg(-R\5 -
\frac14F\5^{\mu\nu}F_{(5)\mu\nu}\bigg) \nn \\ && -
\frac1{12\sqrt3}\epsilon^{\mu\nu\rho\sigma\lambda}F_{(5)\mu\nu}
F_{(5)\rho\sigma}A_{(5)\lambda}\bigg]\,, \ea where $F\5 = dA\5$,
$\mu,\nu,\cdots = 1,\cdots,5$, and
$\epsilon^{\mu\nu\rho\sigma\lambda}$ is the five-dimensional
antisymmetric  symbol. This leads to the five-dimensional
Maxwell-Chern-Simons and  Einstein equations  \ba
\sqrt{|g\5|}D_{\mu}F\5^{\mu\nu} &=& \frac1{4\sqrt3}
\epsilon^{\nu\rho\sigma\tau\lambda}F_{(5)\rho\sigma}F_{(5)\tau\lambda}\,,
\lb{MCS}\\ {R\5^{\mu}}_{\nu} - \frac12R\5\delta^{\mu}_{\nu} &=&
\frac12F\5^{\mu\rho} F_{5\nu\rho}-\frac18F\5^2\delta^{\mu}_{\nu}\,,
\lb{E} \ea  where $D_{\mu}$ is the covariant derivative associated
with the metric  $g_{(5)\mu\nu}$.

Assuming the existence of two Killing vectors, one can choose adapted
coordinates such that the five-dimensional metric and electromagnetic
potential depend only on three coordinates $x_i$ ($i=1,2,3$) and split
according to:  \ba  ds_{(5)}^2 &=& \lambda_{ab}(dz^a + a_i^adx^i)(dz^b
+ a_j^bdx^j) + \tau^{-1}h_{ij}dx^idx^j\,, \\ A_{(5)} &=& \sqrt3(\psi_a
dz^a + A_idx^i)\,,  \ea  where $a,b=0,1$ ($x^4 = z^0,x^5= z^1$), and
$\tau \equiv |det\lambda|$. The reduced metric  $h_{ij}$ or its
inverse $h^{ij}$ will be used to lower or raise indices  $i,j,k$,  and
we will denote by $\nabla_j$ the associated covariant derivative.  The
$\nu = i$ components of the Maxwell-Chern-Simons equations (\ref{MCS})
can be written as \be \nabla_j\bigg(\tau\bigg[F^{ij} +
a^{ai}\partial^j\psi_a -  a^{aj}\partial^i\psi_a\bigg]\bigg) =
\nabla_j\bigg(\frac1{\sqrt{h}}
\epsilon^{ab}\epsilon^{ijk}\psi_a\partial_k\psi_b\bigg)\,, \ee where
$F_{ij} \equiv \partial_iA_j - \partial_jA_i$. This equation  allows
to dualize the vector magnetic potential $A_i$ to a scalar magnetic
potential $\mu$ defined by
\begin{equation}\lb{dualmu}
F^{ij} = a^{aj} \partial^i \psi_a - a^{ai} \partial^j \psi_a +
\frac1{\tau \sqrt{h}} \epsilon^{ijk} \eta_k\,, \qquad \eta_k =
\partial_k \mu +   \epsilon^{ab} \psi_a \partial_k\psi_b \,.
\end{equation}
Similarly, the $\mu=i$, $\nu = a$ components of the Einstein equations
(\ref{E}) read \be \nabla_j\left(\tau\lambda_{ab}G^{bij}\right)  = -
\epsilon^{ijk}\nabla_j\left(\frac1{\sqrt{h}}\psi_a\left[3\partial_k\mu
+ \epsilon^{bc}\psi_b\partial_k\psi_c\right]\right)\,, \ee where
$G^b_{ij} \equiv \partial_ia^b_j - \partial_ja^b_i$. This is
integrated by
\begin{equation}\lb{dualom}
\lambda_{ab}G^{bij} =  \frac1{\tau \sqrt{h}} \epsilon^{ijk} V_{ak}\,,
\qquad V_{ak} =  \partial_k\omega_a - \psi_a\left(3\partial_k\mu +
\epsilon^{bc}  \psi_b\partial_k\psi_c\right)\,,
\end{equation}
where $\omega_a$ is the `twist' or gravimagnetic two-potential
\cite{maison}. It is then straightforward to show that the  $\mu=i$,
$\nu = j$ components of the five-dimensional Einstein  equations
(\ref{E}) lead to the following three-dimensional  Einstein equations:
\ba R_{ij} &=& \frac14 Tr(\lambda^{-1}\partial_i\lambda\lambda^{-1}
\partial_j\lambda) + \frac14\tau^{-2}\partial_i\tau\partial_j\tau  -
\frac12\tau^{-1}V_i^T\lambda^{-1}V_j \nonumber \\ &&  +
\frac32\left(\partial_i\psi^T\lambda^{-1}\partial_j\psi  -
\tau^{-1}\eta_i\eta_j\right) \,, \ea  where $\lambda$ is the
$2\times2$ matrix of elements $\lambda_{ab}$,  $V_i$ the column matrix
of elements $V_{ai}$, and $R_{ij}$ the Ricci  tensor built out of the
reduced metric $h_{ij}$. These equations,  together with the other
field equations arising from the dimensional  reduction of the
original five-dimensional field equations, derive  from the reduced
action (up to a multiplicative constant)
\begin{equation}\lb{sig}
S_3=\int
 d^3x\sqrt{h}\left(-R+\frac12G_{AB}\frac{\partial\Phi^A}{\partial x^i}
 \frac{\partial\Phi^B}{\partial x^j}h^{ij}\right),
\end{equation}
where the $\Phi^A$ ($A=1,\cdots,8$) are the eight scalar fields
$\lambda_{ab}$, $\omega_a$, $\psi_a$, and $\mu$. The action
(\ref{sig})  describes  the three-dimensional gravity coupled sigma
model for  the eight-dimensional target space $\cal T$ with metric:
\ba\lb{tarmet} dS^2 &\equiv& G_{AB}d\Phi^Ad\Phi^B =  \frac12
Tr(\lambda^{-1}d\lambda\lambda^{-1}d\lambda) + \frac12\tau^{-2}d\tau^2
- \tau^{-1}V^T\lambda^{-1}V \nonumber\\ && +
3\left(d\psi^T\lambda^{-1}d\psi - \tau^{-1}\eta^2\right) \,, \ea where
\be \eta = d\mu +   \epsilon^{ab} \psi_a d\psi_b \,, \qquad V_{a} =
d\omega_a - \psi_a\left(3d\mu + \epsilon^{bc}\psi_bd\psi_c\right)\,.
\end{equation}

\section{The symmetry algebra}
\setcounter{equation}{0}

The Killing equations \be\lb{kil} J_{A;B} + J_{B;A} = 0 \ee for the
metric (\ref{tarmet}) constitute a system of 36 partial derivative
equations for eight unknown functions of eight variables. The
analytical   solution of this system is possible in principle, but
represents a  formidable task which is best adressed by computer
\cite{g2}. However  it is possible to find the Killing vectors of
(\ref{tarmet}) without  explicitly solving (\ref{kil}) by using
information on the manifest  symmetries coming from the
field-theoretical construction of (\ref{tarmet}),  combined with
information about the hidden symmetries of five-dimensional  pure
gravity (the vacuum sector of five-dimensional minimal supergravity)
and with the Jacobi identities.

The manifest symmetries of $\cal T$ have two origins. First,  the
original gauge invariances of (\ref{em5}) --- diffeomorphism
invariance for the five-dimensional tensor fields  $g_{(5)\mu\nu}$ and
$A_{(5)\mu}$, and gauge invariance for the gauge  field $A\5$ --- are
broken by the dimensional reduction down to the  corresponding
diffeomorphism invariance for the three-dimensional  tensor, vector
and scalar fields, together with $GL(2,R)$ invariance  (freedom of
choice of  two basis vectors in the two-plane ($z^0,z^1$)) and
residual gauge  invariance for the `electric' potentials
$\psi_a$. Second, the duality  equations (\ref{dualmu}) and
(\ref{dualom}) define the cyclic coordinates  $\mu$ and $\omega_a$
only up to translations.

The corresponding infinitesimal symmetries lead to nine Killing
vectors,  a $GL(2,R)$ tensor, two vectors, and a scalar. The four
components of the mixed tensor \be {M_a}^b =
2\lambda_{ac}\frac{\partial}{\partial\lambda_{cb}} +
\omega_a\frac{\partial}{\partial\omega_{b}} +
\delta_a^b\omega_c\frac{\partial}{\partial\omega_{c}} +
\psi_a\frac{\partial}{\partial\psi_{b}} +
\delta_a^b\mu\frac{\partial}{\partial\mu} \ee generate linear
transformations in the $(z^0,z^1)$ plane obeying the $gl(2,R)$
subalgebra, \be\lb{MM} \left[{M_a}^b,{M_c}^d\right] =
\delta_c^b{M_a}^d - \delta_a^d{M_c}^b\,.  \ee The two-vector and the
scalar associated with the the three cyclic `magnetic' coordinates:
\ba N^a &=& \frac{\partial}{\partial\omega_a}\,, \\ Q &=&
\frac{\partial}{\partial\mu} \ea follow the commutation relations \ba
\left[{M_a}^b,N^c\right] &=& -(\delta_a^cN^b + \delta_a^bN^c)\,,
\lb{MN}\\ \left[{M_a}^b,Q\right] &=& -\delta_a^bQ\,, \lb{MQ}\\
\left[N^a,N^b\right] &=& 0\,,   \lb{NN}\\ \left[Q, N^a\right] &=&
0\,. \lb{QN} \ea Infinitesimal gauge transformations of the $\psi_a$
are generated by the  two-vector  \be R^a =
\frac{\partial}{\partial\psi_a} +
3\mu\frac{\partial}{\partial\omega_a} -
\epsilon^{ab}\psi_b\left(\frac{\partial}{\partial\mu} +
\psi_c\frac{\partial}{\partial\omega_c}\right) \ee  with the
commutation relations  \ba \left[{M_a}^b,R^c\right] &=& -\delta_a^cR^b
\,, \lb{MR}\\ \left[N^a,R^b\right] &=& 0\,, \lb{NR}\\
\left[Q,R^a\right] &=& 3N^a\,, \lb{QR}\\ \left[R^a,R^b\right] &=&
2\epsilon^{ab}Q\,. \lb{RR} \ea

An exact solution of the five-dimensional field equations
(\ref{MCS}), (\ref{E}) is $A\5=0$, corresponding to
five-dimensional Einstein gravity. It is  natural to assume that the
isometry group $SL(3,R)$ \cite{maison} of the  corresponding target
space ((\ref{tarmet}) with $\psi = \mu = 0$) is a  subgroup of the
isometry group $G$ of the full target space. This means  that there must
exist two more `hidden' Killing vectors $L_a$ completing  the
subalgebra $sl(3,R)$: \ba \left[{M_a}^b,L_c\right] &=& (\delta_c^bL_a +
\delta_a^bL_c)\,,\lb{ML}\\ \left[N^a,L_b\right] &=& {M_b}^a\,,
\lb{NL}\\ \left[L_a,L_b\right] &=& 0\,. \lb{LL} \ea Adding to the
known form of the $SL(3,R)$ for five-dimensional Einstein gravity the
information from (\ref{NL}), we know that \ba L_a &=&
\omega_a\omega_b\frac{\partial}{\partial\omega_b} +
2\omega_b\lambda_{ac}\frac{\partial}{\partial\lambda_{bc}} +
\omega_b\psi_a\frac{\partial}{\partial\psi_b} +
\omega_a\mu\frac{\partial}{\partial\mu} \nonumber \\ &&+
\tau\lambda_{ab}\frac{\partial}{\partial\omega_b} + \cdots \ea (the
omitted terms are of order 0 in $\omega_a$). Commutation with $Q$,
\be\lb{QL} \left[Q,L_a\right] = P_a \,, \ee gives two more generators
\be P_a = \omega_a \frac{\partial}{\partial\mu} + \cdots\,, \ee which
now adds up to 13 generators. Finally commutation with the $R^a$
should lead in principle to four more generators, a traceless tensor
${S_a}^b$ and a scalar $T$, \be \left[R^a,L_b\right] = {S_b}^a +
\delta_b^aT\,.  \ee At this stage we make the second, crucial,
assumption that the algebra $Lie(G)$ is minimal and closes with a single
scalar generator $T$ (${S_a}^b=0$): \be\lb{RL} \left[R^a,L_b\right] =
\delta^a_bT \,.  \ee This gives \be T =
\omega_c\frac{\partial}{\partial\psi_c} +
3\mu\omega_c\frac{\partial}{\partial\omega_c} + \cdots\,.  \ee

Now the full algebra can be found using the following three
constraints:

1) Commutators must respect the Jacobi identities.

2) It follows from the Jacobi identities involving the traceless part
of ${M_a}^b$ that the commutators of tensorial operators are
tensors. The only constant tensors are the Kronecker symbol
$\delta^a_b$ and the antisymmetric symbols $\epsilon^{ab}$ and
$\epsilon_{ab}$.

3) It also follows from the Jacobi identities involving the trace
$Tr(M)  \equiv {M_c}^c$ that commutators must respect dimension. The
degrees  (logarithmic dimensions) of the various fields are, in a
scale such that $\omega_a$ has degree 1, \be [\psi_a] = 1/3\,, \quad
[\mu] = [\lambda_{ab}] = 2/3\,, \quad [\omega_a] = 1\,.  \ee This
leads to the degrees of the various Killing vectors \ba
\left[{M_a}^b\right] &=& 0\,, \nonumber \\ \left[P_a\right] =
1/3\,,\qquad && \left[R^a\right] = -1/3\,,  \nonumber \\
\left[T\right] = 2/3\,, \qquad && \left[Q\right] = -2/3\,,   \nonumber
\\ \left[L_a\right] = 1\,, \qquad && \left[N^a\right] = -1\,.    \ea

The full algebra consists of the above commutation relations
(\ref{MM}), (\ref{MN}), (\ref{MQ}), (\ref{NN}), (\ref{QN}),
(\ref{MR}), (\ref{NR}), (\ref{QR}), (\ref{RR}), (\ref{ML}),
(\ref{NL}), (\ref{LL}), (\ref{QL}), (\ref{RL}) together with \ba
\left[{M_a}^b,P_c\right] &=& \delta^b_cP_a \,, \lb{MP}\\
\left[{M_a}^b,T\right] &=& \delta^b_aT \,, \lb{MT}\\
\left[{N^a},P_b\right] &=& \delta^a_bQ \,, \lb{NP}\\
\left[{N^a},T\right] &=& R^a \,, \lb{NT}\\ \left[Q,P_a\right] &=&
-2\epsilon_{ab}R^b\,, \lb{QP}\\ \left[Q,T\right] &=& Tr(M)\,,
\lb{QT}\\ \left[{R^a},P_b\right] &=& -3{M_b}^a + \delta^a_bTr(M)
\,,\lb{RP}\\ \left[{R^a},T\right] &=& 2\epsilon^{ab}P_b \,, \lb{RT}\\
\left[{L_a},P_b\right] &=& 0\,, \lb{LP}\\ \left[{L_a},T\right] &=&
0\,, \lb{LT}\\ \left[{P_a},P_b\right] &=& 2\epsilon_{ab}T\,, \lb{PP}\\
\left[{P_a},T\right] &=& 3L_a \,. \lb{PT} \ea

This is a rank 2 algebra which can be put in the Cartan-Weyl form,
with \be
\begin{array}{lll} H_1 = {\ds\frac{{M_0}^0 + {M_1}^1}{\sqrt6}}\,,
& H_2 = {\ds\frac{{M_0}^0 - {M_1}^1}{\sqrt2}}\,, & \\ E_1 = {M_0}^1\,,
\quad & E_{-1} = {M_1}^0\,, \quad & \alpha_1 = (0,\sqrt2)\,, \\ E_2 =
\frac1{\sqrt3}P_0\,, \quad & E_{-2} = \frac1{\sqrt3}R^0\,,  \quad &
\alpha_2 = ({\ds\frac1{\sqrt6},\frac1{\sqrt2}})\,, \\ E_3 =
\frac1{\sqrt3}P_1\,, \quad & E_{-3} = \frac1{\sqrt3}R^1\,,  \quad &
\alpha_3 = ({\ds\frac1{\sqrt6},\frac{-1}{\sqrt2}})\,, \\ E_4 =
\frac1{\sqrt3}T\,, \quad & E_{-4} = \frac1{\sqrt3}Q\,,  \quad &
\alpha_4 = ({\ds\frac{\sqrt2}{\sqrt3}},0)\,, \\ E_5 = L_0\,, \quad &
E_{-5} = -N^0\,, \quad & \alpha_5 =
({\ds\frac{\sqrt3}{\sqrt2},\frac1{\sqrt2}})\,, \\ E_6 = L_1\,, \quad &
E_{-6} = -N^1\,, \quad & \alpha_6 =
({\ds\frac{\sqrt3}{\sqrt2},\frac{-1}{\sqrt2}})\,,
\end{array}
\ee with $\alpha_4 = \alpha_2+\alpha_3$, $\alpha_1 =
\alpha_2-\alpha_3$, $\alpha_6 = \alpha_3+\alpha_4$, $\alpha_5 =
\alpha_2+\alpha_4 = \alpha_1+\alpha_6$. The root space diagram is that
of the 14-dimensional algebra $Lie(G_2$).  Note that the roots are
arranged in order of increasing degree,  and that the hidden symmetry
generators ($P_a$, $T$, $L_a$) correspond  to the five roots with
positive abscissa.

\section{Determination of the five hidden symmetries}
\setcounter{equation}{0}

Our strategy is to exploit the above commutation relations to
determine the uncompletely known Killing vectors $L_a$, $P_a$ and
$T$. First, the commutation relations (\ref{NT}) and (\ref{QT}) can be
solved to yield \be T = \omega_bR^b + \mu\left[2\lambda_{bc}
\frac{\partial}{\partial\lambda_{cb}} +
\psi_b\frac{\partial}{\partial\psi_{b}} +
\mu\frac{\partial}{\partial\mu}\right] + X\,, \ee where \be
\frac{\partial X}{\partial\omega_{a}} = 0\,, \quad \frac{\partial
X}{\partial\mu} = 0\,.  \ee The unknown $X$ is parametrized by \be X =
X_A\partial_A \equiv X_{\lambda_{cd}}
\frac{\partial}{\partial\lambda_{dc}} +
X_{\omega_{c}}\frac{\partial}{\partial\omega_{c}} +
X_{\psi_c}\frac{\partial}{\partial\psi_{c}} +
X_{\mu}\frac{\partial}{\partial\mu}\,.  \ee Then, relation (\ref{RT})
gives \ba P_a &=&
-3\mu^2\epsilon_{ab}\frac{\partial}{\partial\omega_{b}} -
2\mu\epsilon_{ab}\frac{\partial}{\partial\psi_{b}} -
2\mu\psi_a\left(\frac{\partial}{\partial\mu} + \psi_b
\frac{\partial}{\partial\omega_{b}}\right) \nonumber \\ && +
\omega_a\frac{\partial}{\partial\mu} -
\psi_a\left(\lambda_{bc}\frac{\partial}{\partial\lambda_{cb}}
+\psi_b\frac{\partial}{\partial\psi_{b}}\right)
-\frac12\epsilon_{ab}\left(\frac{\partial
X_A}{\partial\psi_b}\partial_A - X_{\mu}
\frac{\partial}{\partial\omega_{b}}\right)\nonumber\\ &&
+\frac12X_{\psi_a}\left(\frac{\partial}{\partial\mu} + \psi_b
\frac{\partial}{\partial\omega_{b}}\right) + \frac12\psi_aX_{\psi_b}
\frac{\partial}{\partial\omega_{b}}\,.  \ea Inserting this in relation
(\ref{RP}) gives a system of eight second order differential equations
for eight unknown functions $X_A$ of two variables $\psi_c$ (also
depending on the three ``parameters'' $\lambda_{cd})$: \ba
\left[R^a,P_b\right] &=& -3\omega_b\frac{\partial}{\partial\omega_{a}}
-3\psi_b\frac{\partial}{\partial\psi_{a}} - \delta^a_b
\left(\lambda_{cd}\frac{\partial}{\partial\lambda_{dc}} -
\psi_d\frac{\partial}{\partial\psi_{d}} +
\mu\frac{\partial}{\partial\mu}\right) \nonumber\\ && -
\frac12\epsilon_{bc}\left(\frac{\partial^2
X_A}{\partial\psi_a\partial\psi_c}\partial_A - \frac{\partial
X_{\mu}}{\partial\psi_c} \frac{\partial}{\partial\omega_a} -
\frac{\partial X_{\mu}}{\partial\psi_a}
\frac{\partial}{\partial\omega_c}\right) \nonumber \\ &&
+\left(\frac{\partial X_{\psi_b}}{\partial\psi_a}-\frac12
\delta^a_b\frac{\partial X_{\psi_c}}{\partial\psi_c}\right)\left(
\frac{\partial}{\partial\mu} + \psi_d
\frac{\partial}{\partial\omega_{d}}\right) \\&&
+\frac{\partial X_{\psi_c}}{\partial\psi_d}
\left(\delta^a_d\psi_b-\frac12\delta^a_b\psi_d\right)
\frac{\partial}{\partial\omega_{c}} -
\left(X_{\psi_b}\frac{\partial}{\partial\omega_{a}}
-\frac12\delta^a_bX_{\psi_c}\frac{\partial}{\partial\omega_{c}}\right)
\nonumber\\ &=& -6\lambda_{bc}\frac{\partial}{\partial\lambda_{ca}} -
3\omega_b\frac{\partial}{\partial\omega_{a}} -
3\psi_b\frac{\partial}{\partial\psi_{a}} +
\delta^a_b\left(2\lambda_{cd}\frac{\partial}{\partial\lambda_{dc}} +
\psi_c\frac{\partial}{\partial\psi_{c}} -
\mu\frac{\partial}{\partial\mu}\right)\,. \nn
\ea
The only inhomogeneous equation is
\be \frac{\partial^2
X_{\lambda_{cd}}}{\partial\psi_a\partial\psi_b} =
6\epsilon^{be}\left(\delta^a_e\lambda_{cd}-
\delta^a_c\lambda_{de}-\delta^a_d\lambda_{ce}\right)\,.  \ee This is
solved by \be X_{\lambda_{ab}} =
6\epsilon^{cd}\lambda_{(ac}\psi_{b)}\psi_d + G_{ab}\,, \ee where the
symmetric tensor $G_{ab}$ depends only on the $\lambda_{cd}$ (a
component linear in the $\psi_d$ will not lead to a second order
tensor). The only possibility is $G_{ab} = f(\tau)\lambda_{ab}$, and
$[f] = [X] = 2/3$ means that necessarily $f(\tau) \propto \tau^{1/2}$.
Such a fractional power can be reasonably excluded to occur in the
Killings of (\ref{tarmet}), leading to \be G_{ab} = 0\,.  \ee Next, we
turn to the component equation along $\partial/\partial\psi_{b}$, \be
\frac{\partial^2 X_{\psi_b}}{\partial\psi_c\partial\psi_d} = 0\,, \ee
which is solved by \be X_{\psi_b} = {F_b}^c(\lambda)\psi_c \ee (a term
of degree 0 in the $\psi_d$ would not lead to a vector). Only one
mixed tensor of the correct dimension can be constructed from
$\lambda_{cd}$ (without involving fractional powers of $\tau$), this
is ${F_b}^c = \alpha\epsilon^{cd}\lambda_{bd}$ ($\alpha$
constant). Thus, \be X_{\psi_b} =
\alpha\epsilon^{cd}\lambda_{bd}\psi_c\,.  \ee Inserting this into the
component along $\partial/\partial\mu$ and using the identity
\be\lb{ideps1} \epsilon^{ac}\lambda_{bc} =
\tau\epsilon_{bc}\lambda^{ac} \ee (where the $\lambda^{ac}$ are the
elements of the matrix  $\lambda^{-1}$) leads to the equation \be
\frac{\partial^2 X_{\mu}}{\partial\psi_a\partial\psi_b} =
2\alpha\tau\lambda^{ab}\,, \ee which is solved by \be X_{\mu} =
\alpha\tau\lambda^{ab}\psi_a\psi_b + \beta\tau\,, \ee with $\beta$ a
new integration constant. Finally, the component along
$\partial/\partial\omega_c$ reads \be \frac{\partial^2
X_{\omega_c}}{\partial\psi_a\partial\psi_b} =
4\alpha\tau\left[\left(\delta^b_c\lambda^{ad} +
\delta^a_c\lambda^{bd}\right)\psi_d + \lambda^{ab}\psi_c\right]\,, \ee
which is solved by \be X_{\omega_c}=
2\alpha\tau\lambda^{ab}\psi_a\psi_b\psi_c + \gamma\tau\psi_c\,, \ee
with $\gamma$ a third integration constant.

Now, the generators $T$ and $P_a$ are known up to three undetermined
constants: \ba T &=& \left[2\mu\lambda_{bc} +
6\epsilon^{de}\lambda_{bd}\psi_c\psi_e\right]
\frac{\partial}{\partial\lambda_{bc}} \nonumber \\ &+&
\left[3\mu\omega_{b} + \gamma\tau\psi_b -
\epsilon^{cd}\omega_{c}\psi_b\psi_d +
2\alpha\tau\lambda^{cd}\psi_b\psi_c\psi_d\right]
\frac{\partial}{\partial\omega_{b}} \nonumber \\ &+& \left[\omega_b +
\mu\psi_{b} + \alpha\epsilon^{cd}\lambda_{bd}\psi_c\right]
\frac{\partial}{\partial\psi_{b}} \lb{T2} \\ &+& \left[\mu^2 +
\beta\tau - \epsilon^{bc}\omega_{b}\psi_c +
\alpha\tau\lambda^{bc}\psi_b\psi_c\right]
\frac{\partial}{\partial\mu}\,,\nn \ea
\ba P_a &=&
\left[2\lambda_{bc}\psi_a - 6\lambda_{ab}\psi_c\right]
\frac{\partial}{\partial\lambda_{bc}} \nonumber \\ &+&
\left[-3\mu^2\epsilon_{ab} + \frac{3\beta-\gamma}2\tau\epsilon_{ab} -
2\mu\psi_a\psi_b - \alpha\epsilon^{cd}\lambda_{ad}\psi_b\psi_c\right]
\frac{\partial}{\partial\omega_{b}} \nonumber \\ &+&
\left[-2\mu\epsilon_{ab} + \frac{\alpha}2\lambda_{ab} -
\psi_a\psi_b\right] \frac{\partial}{\partial\psi_{b}} \lb{P2}\\ &+&
\left[\omega_a - 2\mu\psi_{a} -
\frac{\alpha}2\epsilon^{cd}\lambda_{ad}\psi_c\right]
\frac{\partial}{\partial\mu}\,.\nn
\ea
The remaining two generators $L_a$ can then be computed from (\ref{PT}):
\ba
L_a &=&
\left[2\omega_b\lambda_{ac} + 2\mu\left(\lambda_{bc}\psi_a -
3\lambda_{ab}\psi_c\right)  +
2\epsilon^{de}\lambda_{bd}\psi_a\psi_c\psi_e\right]
\frac{\partial}{\partial\lambda_{bc}} \nonumber \\ &+&
\left[\omega_a\omega_b + \frac{\alpha\gamma}6\tau\lambda_{ab}
-\mu^3\epsilon_{ab} + \frac{3\beta-\gamma}2\mu\tau\epsilon_{ab} -
\mu^2\psi_a\psi_b \right.\nonumber\\&&\left.  -
\alpha\mu\epsilon^{cd}\lambda_{ad}\psi_b\psi_c
+\frac{8\alpha^2+7(\beta-\gamma)}6\tau\psi_a\psi_b +
\alpha\tau\lambda^{cd}\psi_a\psi_b\psi_c\psi_d
\right]\frac{\partial}{\partial\omega_{b}} \nonumber \\ &+& \left[-
\mu^2\epsilon_{ab} + \frac{\alpha}2\mu\lambda_{ab} +
\frac{7\beta-\gamma-\alpha^2}6\tau\epsilon_{ab} + \omega_b\psi_a
\right.\lb{L2}\\&&\left.  - \mu\psi_a\psi_b +
\frac{\alpha}2\epsilon^{cd}\lambda_{bd}\psi_a\psi_c\right]
\frac{\partial}{\partial\psi_{b}} \nonumber \\ &+& \left[\mu\omega_a -
\mu^2\psi_{a} - \frac{\alpha}2\mu\epsilon^{cd}\lambda_{ad}\psi_c
+\frac{\alpha^2+\beta-\gamma}2\tau\psi_a +
\frac{\alpha}2\tau\lambda^{bc}\psi_a\psi_b\psi_c\right]
\frac{\partial}{\partial\mu}\,,\nn
\ea
consistent with (\ref{ML}) ($L_a$ is a vector of degree 1),  (\ref{NL})
and (\ref{QL}). Computation of the commutator $[R^a,L_b]$ leads to
\ba
\left[R^a,L_b\right] &=& \delta^a_b\bigg\{\left[2\mu\lambda_{cd} +
6\epsilon^{ef}\lambda_{ce}\psi_d\psi_f\right]
\frac{\partial}{\partial\lambda_{cd}} \nonumber \\ &+&
\left[3\mu\omega_{c} + \left(\alpha^2+2\beta-\gamma\right)
\tau\psi_c\right.\nonumber\\ && \left.  -
\epsilon^{de}\omega_{d}\psi_c\psi_e +
2\alpha\tau\lambda^{de}\psi_c\psi_d\psi_e\right]
\frac{\partial}{\partial\omega_{c}} \nonumber \\ &+& \left[\omega_c +
\mu\psi_{c} + \alpha\epsilon^{de}\lambda_{ce}\psi_d\right]
\frac{\partial}{\partial\psi_{c}} \nonumber \\ &+& \left[\mu^2 +
\frac{\alpha^2+5\beta-2\gamma}3\tau \right.\nonumber\\ && \left.  -
\epsilon^{cd}\omega_{c}\psi_d +
\alpha\tau\lambda^{cd}\psi_c\psi_d\right]
\frac{\partial}{\partial\mu}\bigg\}\,, \ea
consistent with (\ref{RL})
provided \be\lb{const1} \alpha^2 + 2(\beta - \gamma) = 0.  \ee Finally
there remains to check the commutation relations (\ref{LL}) (the
remaining commutation relations will then be satisfied by virtue  of
the Bianchi identities). A lengthy computation leads to \ba \left[L_a,
L_b\right] &=& (3\beta-\gamma)\tau\bigg\{
\lambda_{[ac}\psi_{b]}\bigg(8\psi_d\partial_{\lambda_{cd}} -
\frac{2\alpha}3\partial_{\psi_c}\bigg) \nn\\ && -
\epsilon_{ab}\bigg[\frac7{12}\alpha\tau(\alpha - \lambda^{de}
\psi_d\psi_e)+\frac14\mu^2\bigg]\psi_c\partial_{\omega_c} \nn\\ && +
\frac{\alpha}2\tau\epsilon_{ab}\bigg(\frac{\alpha}4+\lambda^{de}
\psi_d\psi_e\bigg)\partial_{\mu} \bigg\}\,, \ea where we have taken
(\ref{const1}) into account. So (\ref{LL}) is satisfied  provided
\be\lb{const2} \beta=\frac{\gamma}3 = \frac{\alpha^2}4\,.  \ee

We have thus obtained a realisation of $Lie(G_2)$ depending on an
arbitrary  real parameter $\alpha$. The reason for this arbitrariness
may be traced to  the fact that the algebra (\ref{MM}),
(\ref{MN})-(\ref{QN}),  (\ref{MR})-(\ref{RR}) of the manifest
symmetries is invariant under the  combined scale transformation
$\Phi^A \to \Phi'^A$ with \be\lb{scale} \lambda'_{ab} =
\lambda_{ab}\,, \quad \psi'_a = k\psi_a\,, \quad \mu' = k^2\mu\,,
\quad \omega'_a = k^3\omega_a\,, \ee depending on a real parameter
$k$. The five hidden symmetries transform  according to their
appropriate scales ($P'_a = kP_a$, $T' = k^2T$, $L'_a =  k^3L_a$)
provided the integration constant $\alpha$ is also rescaled to \be
\alpha' = k^2\alpha\,.  \ee However the target space metric
(\ref{tarmet}) is not invariant under the  transformation
(\ref{scale}): \ba dS'^2 &=&  \frac14
Tr(\lambda'^{-1}d\lambda'\lambda'^{-1}d\lambda') +
\frac14\tau'^{-2}d\tau'^2 - \frac12k^6\tau'^{-1}V'^T\lambda^{-1}V'
\nonumber\\ && + \frac32\left(k^2d\psi'^T\lambda'^{-1}d\psi' -
k^4\tau'^{-1}\eta'^2\right) \,.  \ea  Let us determine the value of
$\alpha$ such that $T$ is a Killing vector of  the target space metric
(\ref{tarmet}) (relations (\ref{RT}) and (\ref{PT})  then imply that
the $L_a$ and the $P_a$ are also Killing vectors). The action of $T$
leads to the first order variations (written in matrix notation) \ba
\delta\lambda &=& 2\mu\lambda + 3(\lambda J\psi\cdot\psi -
\psi\cdot\psi J\lambda)\,, \\ \delta\omega &=& 3\mu\omega +
\frac{3\alpha^2}4\tau\psi - (\omega J\psi)\psi +
2\alpha\tau(\psi\lambda^{-1}\psi)\psi\,, \\ \delta\psi &=& \omega +
\mu\psi - \alpha\lambda J\psi\,, \\ \delta\mu &=& \mu^2 +
\frac{\alpha^2}4\tau - (\omega J\psi) +
\alpha\tau(\psi\lambda^{-1}\psi)\,, \ea with $J^{ab} \equiv
\epsilon^{ab}$. This leads to \ba
\frac14\delta(\lambda^{-1}d\lambda)^2 &=& \tau^{-1}d\tau d\mu +
3\left(\psi\lambda^{-1}d\lambda Jd\psi + d\psi\lambda^{-1}d\lambda
J\psi \right)\,, \ea \ba \frac14\delta(\tau^{-1}d\tau)^2 &=&
2\tau^{-1}d\tau d\mu \,, \ea \ba \frac12\delta(d\psi\lambda^{-1}d\psi)
&=& 3\left(\psi Jd\psi\right)\left(\psi\lambda^{-1}d\psi\right) +
\left(d\omega \lambda^{-1}d\psi\right) \nn\\ && +
d\mu\left(\psi\lambda^{-1}d\psi\right) -
\alpha\left(d\psi\lambda^{-1}d\lambda J\psi\right) \,, \ea \ba
-\frac12\delta(\tau^{-1}\eta^2) &=& -\left[d\mu + \psi
Jd\psi\right]\bigg[\frac{\alpha^2}4\tau^{-1}d\tau  \nn\\ && +
2\tau^{-1}\left(\psi Jd\omega\right) +
2\alpha\left(\psi\lambda^{-1}d\psi\right)\bigg]\,, \ea \ba
-\frac12\delta(\tau^{-1}V\lambda^{-1}V) &=& \tau^{-1}\left[3d\mu +
\left(\psi Jd\psi\right)\right]\bigg[\alpha \left(\psi Jd\omega\right)
+ \frac{3\alpha^2}4\tau\left(\psi\lambda^{-1}d\psi\right)\bigg] \nn\\
&& - \frac{3\alpha^2}4 \left(d\psi\lambda^{-1}d\omega\right)  +
2\alpha\left[\left(\psi\lambda^{-1}d\psi\right)
\left(\psi\lambda^{-1}d\omega\right) \right. \nonumber\\ && \left. -
\left(\psi\lambda^{-1}\psi\right)
\left(d\psi\lambda^{-1}d\omega\right)\right]\,.  \ea Collecting these,
we obtain \ba \delta(dS^2) &=&
3\bigg(1-\frac{\alpha^2}4\bigg)\tau^{-1}d\tau\bigg[d\mu + \bigg(\psi
Jd\psi\bigg)\bigg] +
3\bigg(1-\frac{\alpha^2}4\bigg)\bigg(d\omega\lambda^{-1}d\psi\bigg)
\nonumber\\ && +3(\alpha-2)\tau^{-1}\bigg[d\mu + \bigg(\psi
Jd\psi\bigg)\bigg] \bigg(\psi Jd\omega\bigg) \nonumber\\ && +
3(2-\alpha)\bigg(d\psi\lambda^{-1}d\lambda J\psi\bigg) +
3\bigg(1-2\alpha +\frac{3\alpha^2}4\bigg)d\mu
\bigg(\psi\lambda^{-1}d\psi\bigg) \nonumber\\ &&  +
3\bigg(3-2\alpha+\frac{\alpha^2}4\bigg)\bigg(\psi Jd\psi\bigg)
\bigg(\psi\lambda^{-1}d\psi\bigg) \nonumber\\ && +
3\bigg[\bigg(\psi\lambda^{-1}d\lambda Jd\psi\bigg) -
\bigg(d\psi\lambda^{-1}d\lambda J\psi\bigg) -\tau^{-1}d\tau \bigg(\psi
Jd\psi\bigg)\bigg] \nonumber\\
&&+2\alpha\bigg[\bigg(\psi\lambda^{-1}d\psi\bigg)
\bigg(\psi\lambda^{-1}d\omega\bigg) - \bigg(\psi\lambda^{-1}\psi\bigg)
\bigg(d\psi\lambda^{-1}d\omega\bigg) \nonumber\\ && -
\tau^{-1}\bigg(\psi Jd\psi\bigg) \bigg(\psi Jd\omega\bigg)\bigg]\,.
\ea The last two terms in square brackets vanish identically, while
the remaining terms vanish provided \be \alpha = 2\,, \quad \beta =
1\,, \quad \gamma = 3\,.  \ee

\section{Coset representative}
\setcounter{equation}{0}

In this section we assume for definiteness that the original
five-dimensional metric $g_{(5)\mu\nu}$ is Lorentzian, one of the two
Killing vectors of the original five-dimensional theory (\ref{em5})
being timelike (the corresponding solutions are stationary) and the
other being spacelike. The reduced metric $h_{ij}$ is then Euclidean
and the matrix field $\lambda$ has signature ($+ -$). In the vacuum
sector  ($\psi_a = 0$,  $\mu = 0$), the target space metric (2.12)
reduces to that of the  five-dimensional symmetric space
$SL(3,R)/SL(2,R)$ \cite{maison}, with  signature ($+ + + - -$). The
full eight-dimensional metric (\ref{tarmet}), with signature ($ + + +
+ - - - -$), is that of the symmetric space  $G_{2(2)}/((SL(2,R)\times
SL(2,R))$. This coset was constructed in \cite{mizoh} in terms of the
14-dimensional adjoint representation of $G_{2(2)}$. We present here a
more convenient representation  (previously published without details
in \cite{g2}) in terms of symmetrical $7\times7$ matrices.

The matrix representatives $j_M$ ($M = 1, \cdots, 14$) of the real
form of $Lie(G_2)$ may be derived from the $Z$ matrices of \cite{GG}
by omitting $i$'s.  Their generic block decomposition is \be\lb{jgen}
j = \left(\begin{array}{ccc} S & \tilde{V} & \sqrt2U \\ -\tilde{U} & -
S^T & \sqrt2V \\ \sqrt2V^T &  \sqrt2U^T & 0
\end{array}\right),
\ee where $S$ is a $3\times3$ matrix, $U$ and $V$ are 3-component
column matrices, $U^T$ and $V^T$ the corresponding transposed row
matrices, and $\tilde{U}$, $\tilde{V}$ are the $3\times3$ dual
matrices $\tilde{U}_{ij} = \epsilon_{ijk}U_k$. The matrices ${m_a}^b$,
$n^a$ and $\ell_a$ generating $SL(3,R)$ are of type $S$, the
corresponding $3\times3$ blocks being \ba S_{{m_0}^0}\!\!  &=&\!\!
\left(\begin{array}{ccc}1&0&0\\0&0&0\\0&0&-1
\end{array}\right),\;
S_{{m_0}^1} =
\left(\begin{array}{ccc}0&1&0\\0&0&0\\0&0&0\end{array}\right) ,\;\nn\\
S_{{m_1}^0} &=&
\left(\begin{array}{ccc}0&0&0\\1&0&0\\0&0&0\end{array}\right) ,\;
S_{{m_1}^1} =
\left(\begin{array}{ccc}0&0&0\\0&1&0\\0&0&-1\end{array}\right) ,\;\nn
\\ S_{n^0}\! \!&=&\!\!
\left(\begin{array}{ccc}0&0&0\\0&0&0\\-1&0&0
\end{array}\right),\;
S_{n^1} =
\left(\begin{array}{ccc}0&0&0\\0&0&0\\0&-1&0\end{array}\right),\;\\
S_{\ell_0} &=&
\left(\begin{array}{ccc}0&0&1\\0&0&0\\0&0&0\end{array}\right),\;
S_{\ell_1} =
\left(\begin{array}{ccc}0&0&0\\0&0&1\\0&0&0\end{array}\right).\nn \ea The
matrices $p_a$ and $q$ are of type $U$, the corresponding $1\times3$
blocks being \be U_{p_0} =  \left(\begin{array}{c} 1 \\ 0 \\ 0
\end{array}\right),\;\;
U_{p_1} =  \left(\begin{array}{c} 0 \\ 1 \\ 0
\end{array}\right),\;\;
U_{q} =   \left(\begin{array}{c} 0 \\ 0 \\ -1
\end{array}\right).
\ee The matrices $r^a$ and $t$ are of type $V$,  the corresponding
$1\times3$ blocks being \be V_{r^0} =  \left(\begin{array}{c} 1 \\ 0
\\ 0
\end{array}\right),\;\;
V_{r^1} =  \left(\begin{array}{c} 0 \\ 1 \\ 0
\end{array}\right),\;\;
V_{t} =   \left(\begin{array}{c} 0 \\ 0 \\ 1
\end{array}\right).
\ee  Due to the form of (\ref{jgen}), the transposed matrices $j_A^T$
are related to the original matrices $j_A$ by \be\lb{jtk} j_A^T =
-Kj_AK\,, \ee where the involution $K$ has the block structure \be
\left(\begin{array}{ccc}0&1&0\\1&0&0\\0&0&-1\end{array}\right)\,.  \ee

A representative $N(\Phi^A)$ of the coset $G/H$ (here $G = G_{2(2)}$
and $H = SL(2,R)\times SL(2,R)$) transforms by global right action of
$G$ and local right action of $H$:
\begin{equation}\label{Ntrans}
N(\Phi) \to h(\Phi)N(\Phi)g
\end{equation}
($g \in G$, $h(\Phi) \in H$). The corresponding infinitesimal
transformation is
\begin{equation}\label{Nlocal}
J_M N(\Phi)= N(\Phi)j_M + q_M^\alpha(\Phi)k_{\alpha}N(\Phi)\,,
\end{equation}
where $J_M,\, M=1,\cdots,14$ are Killing vectors acting as
differential operators, $j_M$ are the corresponding matrices of the
$Lie(G)$ algebra,  $y_\alpha,\, \alpha=1,\cdots,6$ are generators of
the isotropy subalgebra $Lie(H)$ and $q_M^\alpha(\Phi)$ are gauge
functions. The gauge can be fixed so that for a suitably chosen Borel
subalgebra $M=A=1,\cdots,8$ the functions  $q_A^\alpha(\Phi)$ vanish,
and the corresponding subset of the equations  (\ref{Nlocal}) reduces
to  \be\lb{Nfix} J_A N(\Phi)= N(\Phi)j_A\,.
\end{equation}
From a solution $N(\Phi)$ of (\ref{Nfix}), one then constructs the
gauge-independent symmetrical matrix , \be  M = N^T\eta N\,, \ee where
$\eta$ is a constant symmetrical matrix invariant under the isotropy
subgroup $H$, \be \eta^T = \eta\,, \quad  h^T\eta h = \eta\,.  \ee The
matrix $M(\Phi)$ is invariant under the local action of $H$ and
transforms  tensorially under the global action of $G$,
\be\lb{globalM}
M(\Phi)\to g^TM(\Phi)g\,.
\ee
The $\sigma$-model current \be\lb{cur} {\cal
J} = M^{-1}dM \ee constructed from the coset representative $M(\Phi)$
is invariant under  the action of $G$. The target space metric
(\ref{tarmet}) is given in terms  of this current by \be dS^2 =
\frac14Tr({\cal J}^2)\,.  \ee Consequently, the current ${\cal J}$ is
conserved by virtue of the field  equations deriving from the
gravitating $\sigma$-model action (\ref{sig})  \be
\frac1{\sqrt{|h|}}\partial_i\bigg(\sqrt{|h|}h^{ij}{\cal J}_j\bigg) =
0\,.  \ee Note that the definition of the matrix $M(\Phi)$ is not
unique, as the current (\ref{cur}) is invariant under $M(\Phi) \to
PM(\Phi) \; (P \in G)$. For  instance a group equivalent coset
representative is \be M' \equiv KM = N^{-1}\eta'N \qquad (\eta' =
K\eta = \eta K)\,, \ee using $N^T = KN^{-1}K$ which follows from
exponentiating (\ref{jtk}).

It is convenient to choose as generators of the Borel subalgebra eight
of the manifest symmetry generators, i.e. three independent components
of  ${M_a}^b$ together with the two $N^a$, $Q$ and the two $R^a$. A
covariant  solution of the first three equations (\ref{Nfix}) would
involve trading the two-metric $\lambda_{ab}$ for a zweibein
${\epsilon_a}^i$. We shall bypass  this by noting that, due to the
structure (\ref{jgen}) of the matrix  generators, the vacuum ($\psi =
\mu = 0$) matrix $M$ is of the form \be M_1 = \left(\begin{array}{ccc}
\chi & 0 & 0 \\ 0 & \chi^{-1} & 0 \\  0 & 0 & 1 \end{array}\right)\,,
\ee with $\chi$ the $SL(3,R)/SL(2,R)$ coset
representative\cite{maison} \be\lb{maimat} \chi =
\left(\begin{array}{cc} \lambda - \tau^{-1}\omega\omega^T &
\tau^{-1}\omega \\ \tau^{-1}\omega^T & -\tau^{-1}
\end{array}\right)\,, \ee where $\lambda$ is a $2\times2$ block, and
$\omega$ a 2-component column  matrix. Thus the static ($\omega = 0$)
vacuum matrix $M$ is \be\lb{M0}  M_0(\lambda) = N(\epsilon)^T\eta
N(\epsilon) = \left(\begin{array}{ccccc} \lambda & 0 & 0 & 0 & 0 \\ 0
& -\tau^{-1} & 0 & 0 & 0 \\ 0 & 0 & \lambda^{-1} & 0 & 0 \\ 0 & 0 & 0
& -\tau & 0 \\ 0 & 0 & 0 & 0 & 1
\end{array}\right)\,.
\ee

The next three equations  \be \frac{\partial}{\partial\omega_a}N =
Nn_a\,, \quad \frac{\partial}{\partial\mu}N = Nq \ee are readily
integrated to: \be\lb{Nmuom} N(\epsilon,\psi,\mu,\omega) =
N(\epsilon,\psi)e^{\mu q}e^{\omega_an^a}\,.  \ee The exponentials in
(\ref{Nmuom}) are easily computed as the matrices  $\mu q$ and $\omega
n\equiv \omega_an^a$ are nilpotent, \be\lb{emuqomn} e^{\mu q} =
\left(\begin{array}{ccccc} 1 & 0 & 0 & 0 & 0 \\ 0 & 1 & 0 & \mu^2 &
-\sqrt2\mu \\ \mu J & 0 & 1 & 0 & 0 \\ 0 & 0 & 0 & 1 & 0 \\ 0 & 0 & 0
& -\sqrt2\mu & 1 \\
\end{array}\right),
\quad e^{\omega n} = \left(\begin{array}{ccccc} 1 & 0 & 0 & 0 & 0 \\
  -\omega^T & 1 & 0 & 0 & 0 \\ 0 & 0 & 1 & \omega & 0 \\ 0 & 0 & 0 & 1
  & 0 \\ 0 & 0 & 0 & 0 & 1 \\
\end{array}\right),
\ee where again the first and third rows and columns are double, and
\be J = \left(\begin{array}{cc} 0 & 1 \\ -1 & 0 \end{array}\right).
\ee

Let us now show that the last two equations \be R^aN = Nr^a \ee are
solved by \be N(\Phi) = N(\epsilon)e^{\psi_ar^a}e^{\mu
q}e^{\omega_an^a}\,.  \ee Again, the matrix $\psi_a r^a$ is nilpotent,
so that \be\lb{epsir} e^{\psi r} = \left(\begin{array}{ccccc} 1 & 0 &
0 & -J\psi & 0 \\ 0 & 1 & -\psi^TJ & 0 & 0 \\ \psi\psi^T & 0 & 1 & 0 &
\sqrt2\psi \\ 0 & 0 & 0 & 1 & 0 \\ \sqrt2\psi^T & 0 & 0 & 0 & 1 \\
\end{array}\right).
\ee This can be used to show that \be \left[R^a,e^{\psi r}\right] =
\frac{\partial}{\partial\psi_a}(e^{\psi r}) =  e^{\psi r}\left[r^a +
\epsilon^{ab}\psi_b(q + \psi_cn^c)\right]\,.  \ee Using this, together
with the commutator, \be \left[r^a,e^{\mu q}\right] = -3e^{\mu q}\mu
n^a\,, \ee we obtain successively \ba R^a e^{\psi r}e^{\mu q}e^{\omega
n} & = &  e^{\psi r}\bigg[r^a + \epsilon^{ab}\psi_b(q + \psi_cn^c) +
R^a\bigg]  e^{\mu q}e^{\omega n} \nn\\ &=&  e^{\psi r}\bigg[r^a + 3\mu
N^a - \epsilon^{ab}\psi_b(Q-q +  \psi_c(N^c-n^c))\bigg] e^{\mu
q}e^{\omega n} \nn\\ &=& e^{\psi r}e^{\mu q}\bigg[r^a + 3\mu(N^a-n^a)
- \epsilon^{ab}\psi_b  \psi_c(N^c-n^c)\bigg] e^{\omega n} \nn\\ &=&
e^{\psi r}e^{\mu q}e^{\omega n}r^a\,.  \ea

The final result for the matrix $M$ is \be\lb{fullM} M(\Phi) =
e^{\omega n^T}e^{\mu q^T}e^{\psi r^T}M_0(\lambda)e^{\psi r} e^{\mu
q}e^{\omega n}.  \ee It follows from (\ref{jgen}) and (\ref{M0}) that
the matrix $M$ has the  symmetrical block structure \be\lb{Mblock} M =
\left(\begin{array}{ccc} A & B & \sqrt2U \\ B^T & C & \sqrt2V \\
\sqrt2U^T & \sqrt2V^T & S
\end{array}\right),
\ee where $A$ and $C$ are symmetrical $3\times3$ matrices, $B$ is a
$3\times3$ matrix, $U$ and $V$ are 3-component column matrices, and
$S$ a scalar. It also follows from (\ref{jtk}) that the inverse matrix
is given by  \be\lb{invmat} M^{-1} = KMK = \left(\begin{array}{ccc} C
& B^T & -\sqrt2V \\ B & A & -\sqrt2U \\ -\sqrt2V^T & -\sqrt2U^T & S
\end{array}\right),
\ee Computation of the product (\ref{fullM}), with the matrices
(\ref{M0}),  (\ref{emuqomn}) and (\ref{epsir}) gives the coset matrix
$M$ in the form (\ref{Mblock}), with
$$
\begin{array}{l}
A = \left(\begin{array}{cc} \begin{array}{c} \left[(1-y)\lambda +
(2+x)\psi\psi^T  - \tau^{-1}\tom\tom^T\right.\\
\left.+\mu(\psi\psi^T\lambda^{-1}J - J\lambda^{-1}\psi\psi^T)\right]
\end{array} & \tau^{-1}\tom \\
\tau^{-1}\tom^T & -\tau^{-1}
\end{array}\right), \\
B = \left(\begin{array}{cc} (\psi\psi^T-\mu J)\lambda^{-1} -
\tau^{-1}\tom\psi^TJ &
\begin{array}{c}
\left[(-(1+y)\lambda J - (2+x)\mu + \psi^T\lambda^{-1}\tom)\psi\right.
\\ \left. + (z - \mu J\lambda^{-1})\tom \right]
\end{array} \\
 \tau^{-1}\psi^TJ & -z
\end{array}\right), \\
C = \left(\begin{array}{cc} (1+x)\lambda^{-1} -
\lambda^{-1}\psi\psi^T\lambda^{-1} & \lambda^{-1}\tom-J(z-\mu
J\lambda^{-1})\psi\\ \tom^T\lambda^{-1} + \psi^T(z+\mu\lambda^{-1}J)J &
\begin{array}{c}
\left[\tom^T\lambda^{-1}\tom - 2\mu\psi^T\lambda^{-1}\tom\right. \\
\left. -\tau(1+x-2y-xy+z^2) \right]
\end{array}
\end{array}\right),
\end{array}
$$\be
\begin{array}{l}
U = \left(\begin{array}{c} (1+x-\mu J\lambda^{-1})\psi -
\mu\tau^{-1}\tom \\ \mu\tau^{-1}
\end{array}\right), \\
V = \left(\begin{array}{c} (\lambda^{-1} + \mu\tau^{-1}J)\psi \\
\psi^T\lambda^{-1}\tom - \mu(1+x-z)
\end{array}\right), \\
S = 1+2(x-y)\,,
\end{array} \lb{coset}
\ee with \be \tom = \omega - \mu\psi\,. \quad x =
\psi^T\lambda^{-1}\psi\,,  \quad y = \tau^{-1}\mu^2\,, \quad z = y -
\tau^{-1}\psi^TJ\tom\,.  \ee

\section{Conclusion}
We have shown that the isometry algebra of the target space for
five-dimensional supergravity reduced to three dimensions is that of
$G_2$ by combining knowledge about the manifest symmetries (gauge
invariances) of the theory with the $SL(3,R)$ invariance of the
vacuum sector. We have then solved the Lie brackets to determine the
generators of the hidden symmetries in terms of the field variables,
and constructed a symmetrical $7\times7$ matrix representative of
the coset $G_{2(+2)}/((SL(2,R)\times SL(2,R))$ as a function of the
same fields.

This coset representative was used in \cite{g2} to generate a
doubly-rotating charged black ring through the action of a group
transformation (\ref{globalM}) on a neutral 5$D$ black ring with two
angular momenta \cite{pomsen, MTY}. After completion of the present
work, several related papers appeared. The geometry of the symmetric
space $G_{2(2)}/SO(4)$ was studied in great  detail in \cite{GNPP},
where in particular the Iwasawa parametrization of the coset and the
Killing vectors were also given explicitly. This approach was
applied in \cite{BP} to the analysis of the supersymmetry
constraints associated with a number of black hole solutions to
gauged and ungauged 5$D$ supergravity. BPS and non-BPS
multi-centered attractor flows were constructed in \cite{GLP},
following the procedure advocated in \cite{multi}.

These works certainly do not exhaust the potentialities of the sigma-model
approach for generating solutions of five-dimensional supergravity. In
\cite{kerr}, it was shown that stationary solutions to the four-dimensional
Einstein-Maxwell equations can be generated from static solutions by a
combination of $SU(2,1)$ group transformations and global coordinate
transformations. This procedure can be extended to generate new stationary
solutions of five-dimensional supergravity, which contains a four-dimensional
Einstein-Maxwell sector \cite{spinem5}. In unrelated recent work, $SL(3,R)$
transformations were also used to generate stationary solutions of
five-dimensional gravity from static solutions \cite{giusax}, and to construct
a static black ring with Kaluza-Klein monopole charge \cite{ford}. It would be
interesting to extend these techniques to the case of five-dimensional
supergravity.

\subsection*{Acknowledgments}
This work was motivated and enriched by stimulating discussions and
exchanges with Dmitry Gal'tsov. I also wish to extend my warm thanks to Paul
Sorba for his continued encouragement and enlightening advice.

\renewcommand{\theequation}{A.\arabic{equation}}
\setcounter{equation}{0}
\section*{Appendix}

We first constructed the coset representative matrix $M(\Phi)$ using
a different procedure. Dualise the Killing vectors ${J_M} =
{J_M}^A\partial/\partial\Phi^A$ to the Killing one-forms \be
\bar{J}_M = G_{AB}{J_M}^Ad\Phi^B\,, \ee where $G_{AB}$ is the target
space metric (\ref{tarmet}). The matrix current  (\ref{cur}) is
proportional to the Killing product of these one-forms with the
matrices $j_M$: \be \lb{J} M^{-1}dM = {\cal J} \equiv 8\sum_M
\eta^{MN}\bar{J}_Mj_N\,, \ee where $\eta^{MN}$ is the inverse of the
Killing metric $\eta_{MN} =  4Tr(j_Mj_N)$. In the present case, the
current (\ref{J}) is given by \ba {\cal J} &=& \left({\bar{M}_b}^a -
\frac13\delta_b^aTr\bar{M}\right){m_a}^b + \bar{N}^an^{aT} +
\bar{L}_a{\ell_a}^T \nonumber \\ && + \frac13\bigg(\bar{R}^ar^{aT} +
\bar{P}_a{p_a}^T + \bar{Q}q^T +  \bar{T}t^T \bigg)\,. \ea The target
space metric can be read off from (\ref{tarmet}): \ba
G_{\lambda_{ac}\lambda_{bd}} & = &
\frac12(\lambda^{(ab}\lambda^{c)d} + \lambda^{ac}\lambda^{bd})\,,\nn
\\ G_{\omega_a\omega_b} & = & -\tau^{-1}\lambda^{ab}\,, \nn\\
G_{\omega_a\mu} &=& 3\tau^{-1}\lambda^{ab}\psi_b \nn\,, \\
G_{\mu\mu} &=& -3\tau^{-1}(1+3\lambda^{ab}\psi_a\psi_b) \,, \\
G_{\omega_a\psi_b} &=&
-\tau^{-1}\lambda^{ac}\epsilon^{bd}\psi_c\psi_d \,,\nn \\
G_{\mu\psi_a}
&=& 3\tau^{-1} (1+\lambda^{cd}\psi_c\psi_d)\epsilon^{ab}\psi_b \,,\nn \\
G_{\psi_a\psi_b} &=& 3\bigg[\lambda^{ab} - \tau^{-1}
(1+\frac13\lambda^{cd}\psi_c\psi_d)\epsilon^{ae}\epsilon^{bf}\psi_e\psi_f
\bigg]\,. \nn
\ea
This leads to the Killing one-forms:
\ba {\bar{M}_a}^b &=&
^b\bigg(\lambda^{-1}d\lambda + \tau^{-1}d\tau -
\tau^{-1}\bigg[\lambda^{-1}d\omega\omega^T +
\omega^T\lambda^{-1}d\omega \nonumber\\&&
-(3\mu-J\psi\psi^T)\psi^T\lambda^{-1}d\omega\bigg]
+3\tau^{-1}\bigg[\lambda^{-1}\psi\omega^T+\omega^T\lambda^{-1}\psi
\nonumber\\&& -\mu(1+3\psi^T\lambda^{-1}\psi) + J\psi\psi^T
(1+\psi^T\lambda^{-1}\psi)\bigg]d\mu \nonumber\\&&
+3\lambda^{-1}d\psi\psi^T+\tau^{-1}\bigg[
(\lambda^{-1}\psi\omega^T+\omega^T\lambda^{-1}\psi) \nonumber\\&& - 3
\mu(1+\psi^T\lambda^{-1}\psi) + 3J\psi\psi^T
(1+\frac13\psi^T\lambda^{-1}\psi)\bigg]\psi^TJd\psi\bigg)_a\,, \nn\\
\bar{N}^a &=& ^a\bigg(-\tau^{-1}\lambda^{-1}\bigg[d\omega -\psi(3d\mu
+ \psi^TJd\psi)\bigg]\bigg)\,,\\
\bar{Q} &=& 3\tau^{-1}
\bigg[-(1+3\psi^T\lambda^{-1}\psi)d\mu + \psi^{T}\lambda^{-1}d\omega -
(1+\psi^T\lambda^{-1}\psi)\psi^TJd\psi\bigg]\,,\nn\\
\bar{R^a} &=&
^a\bigg(3\lambda^{-1}\bigg[d\psi -\tau^{-1}\mu (d\omega -\psi(3d\mu +
\psi^TJd\psi))\bigg]\nonumber\\ &&
+3\tau^{-1}J\psi\bigg[(2+3\psi^T\lambda^{-1}\psi) d\mu -
\psi^{T}\lambda^{-1}d\omega \nonumber \\ && \qquad +
3(2+\psi^T\lambda^{-1}\psi)\psi^TJd\psi\bigg]\bigg) \nn\ea (the last five
Killing one-forms will not be used in the following).

We solved the system of partial differential equations (\ref{J}) in
special cases. In the vacuum sector ($\mu = \psi = 0$), the
symmetrical  solution is the Maison matrix (\ref{maimat}) \be
M_1(\lambda,\omega) = e^{\omega n^T}M_0(\lambda)e^{\omega n}\,.  \ee
In the magnetostatic sector ($\omega = \psi = 0$), the symmetrical
solution is
\be M_2(\lambda,\mu) = e^{\mu q^T}M_0(\lambda)e^{\mu q}\,.
\ee We only give here details on the solution in the electrostatic
sector  ($\omega = \mu = 0$). The symmetrical matrix $M_3$ solves the
equation \be\lb{J3} M_3^{-1}dM_3 = {\cal J}_3, \ee where ${\cal J}_3$
is obtained from ${\cal J}$ by setting $\omega$ and $\mu$ as well as
$d\omega$ and $d\mu$ to zero. This equation constrains the tensorial
characters and degrees of the various matrix elements of $M$ to be (in
$5\times5$ notation) \be M =  \left(\begin{array}{ccccc} M_{ab} &
M_{a3} & {M_a}^b & {M_a}^3 & M_a \\ M_{3b} & M_{33} & {M_3}^b &
{M_3}^3 & M_3 \\ {M^a}_b & {M^a}_3 & M^{ab} & M^{a3} & M^a \\ {M^3}_b
& {M^3}_3 & M^{3b} & M^{33} & M^3 \\ M_b & M_3 & M^b & M^3 & M
\end{array}\right)
\ee (there should be no confusion between these matrix elements and
the Killing vectors ${M_a}^b$), and \be \left[M\right] =
\left(\begin{array}{ccccc} 2/3 & -1/3 & 0 & 1 & 1/3 \\ -1/3 & -4/3 &
-1 & 0 & -2/3 \\ 0 & -1 & -2/3 & 1/3 & -1/3 \\ 1 & 0 & 1/3 & 4/3 & 2/3
\\ 1/3 & -2/3 & -1/3 & 2/3 & 0
\end{array}\right).
\ee These in turn severely constrain the possible dependence of the
matrix elements which must be built from the fields $\lambda_{ab}$,
$\psi_a$, $\tau = \det\lambda$ (recall $[\lambda] = 2/3$ and $[\psi] =
1/3$), and the constant tensors $\epsilon_{ab}$, $\delta_a^b$ and
$\epsilon^{ab}$, with dimensionless  coefficients depending on the
only dimensionless scalar $x \equiv  \psi^T\lambda^{-1}\psi$.

Combining this information with the constraint that the matrix $M_3$
and its inverse are related by (\ref{invmat}), we can determine this
matrix by solving only part of the equations (\ref{J3}). The following
matrix solves the equations (\ref{J3}) for the components ${m_a}^b$,
$n^{aT}$, $r^{aT}$ and $q^T$: \be M_3 = \left(\begin{array}{ccccc}
\lambda + (2+x)\psi\psi^T & 0 & \psi\psi^T\lambda^{-1} & -\lambda
J\psi & \sqrt2(1+x)\psi \\ 0 & -\tau^{-1} & \tau^{-1}\psi^TJ & 0 & 0
\\ \lambda^{-1}\psi\psi^T & -\tau^{-1}J\psi & (1+x)\lambda^{-1} -
\lambda^{-1}\psi\psi^T\lambda^{-1} & 0 & \sqrt2\lambda^{-1}\psi \\
\psi^TJ\lambda & 0 & 0 & -(1+x)\tau & 0 \\ \sqrt2(1+x)\psi^T & 0 &
\sqrt2\psi^T\lambda^{-1} & 0 & 1+2x
\end{array}\right).
\ee
This is of the form (\ref{Mblock}) with the blocks given by
(\ref{coset}), and can be split up as the product
\be\lb{M3M0}
M_3(\lambda,\psi) =  e^{\psi r^T}M_0(\lambda)e^{\psi r},  \ee where
the matrix $e^{\psi r}$ is given in (\ref{epsir}).

\end{document}